 \documentclass[manuscript]{acmart}
\AtBeginDocument{%
  \providecommand\BibTeX{{%
    \normalfont B\kern-0.5em{\scshape i\kern-0.25em b}\kern-0.8em\TeX}}}

\copyrightyear{2025}
\acmYear{2025}
\setcopyright{rightsretained}
\acmConference[CHI EA '25]{Extended Abstracts of the CHI Conference on Human Factors in Computing Systems}{April 26-May 1, 2025}{Yokohama, Japan}
\acmBooktitle{Extended Abstracts of the CHI Conference on Human Factors in Computing Systems (CHI EA '25), April 26-May 1, 2025, Yokohama, Japan}\acmDOI{10.1145/3706599.3720172}
\acmISBN{979-8-4007-1395-8/2025/04}

\begin{document}



\title{OSCAR: Object Status and Contextual Awareness for Recipes to Support Non-Visual Cooking}



\author{Franklin Mingzhe Li}
\affiliation{%
  \institution{Carnegie Mellon University}
  \city{Pittsburgh}
  \state{PA}
  \country{United States}
}
\email{mingzhe2@cs.cmu.edu}

\author{Kaitlyn Ng}
\affiliation{%
  \institution{Carnegie Mellon University}
  \city{Pittsburgh}
  \state{PA}
  \country{United States}
}
\email{kgn@andrew.cmu.edu}

\author{Bin Zhu}
\affiliation{%
  \institution{Singapore Management University}
  \city{Singapore}
  \country{Singapore}
}
\email{binzhu@smu.edu.sg}

\author{Patrick Carrington}
\affiliation{%
  \institution{Carnegie Mellon University}
  \city{Pittsburgh}
  \state{PA}
  \country{United States}
}
\email{pcarrington@cmu.edu}

\renewcommand{\shortauthors}{Li et al.}



\begin{abstract}

Following recipes while cooking is an important but difficult task for visually impaired individuals. We developed OSCAR (Object Status Context Awareness for Recipes), a novel approach that provides recipe progress tracking and context-aware feedback on the completion of cooking tasks through tracking object statuses. OSCAR leverages both Large-Language Models (LLMs) and Vision-Language Models (VLMs) to manipulate recipe steps, extract object status information, align visual frames with object status, and provide cooking progress tracking log. We evaluated OSCAR's recipe following functionality using 173 YouTube cooking videos and 12 real-world non-visual cooking videos to demonstrate OSCAR’s capability to track cooking steps and provide contextual guidance. Our results highlight the effectiveness of using object status to improve performance compared to baseline by over 20\% across different VLMs, and we present factors that impact prediction performance. Furthermore, we contribute a dataset of real-world non-visual cooking videos with step annotations as an evaluation benchmark.

\end{abstract}




\maketitle

\section{Introduction and Background}
Cooking is a fundamental activity that contributes to autonomy and quality of life \cite{li2021choose,li2022freedom}. However, it presents significant barriers for people with vision impairments due to its highly visual and dynamic nature, leading to frustration and increased reliance on pre-packaged foods, which negatively impacts health and nutrition \cite{jones2019analysis, bilyk2009food, kostyra2017food, li2024recipe, wang2023practices,lee2024cookar}. For individuals with vision impairments, following recipes is particularly challenging due to difficulties in tracking progress, accessing future steps, and dealing with limited contextual support provided by existing tools like tactile audio readers and smart speakers \cite{li2024recipe}. These tools often require extensive physical interaction or verbal input, lacking real-time contextual awareness and effective feedback mechanisms during the cooking process \cite{li2024recipe, li2021non}.

Moreover, prior research on tracking instructional tasks primarily utilizes vision models for object detection and action tracking, but in the context of cooking, objects are dynamically transformed, making tracking particularly challenging \cite{lin2020recipe, cao2019video, gong2021temporal, jiao2021new, ciaparrone2020deep, yao2020video, xue2024learning}. For example, ingredients such as an onion may be peeled, chopped, and fried, resulting in significant changes in appearance and form. This transformation necessitates an approach that can accurately recognize and understand these changes to provide effective assistance.

To address these challenges, we developed OSCAR (Object Status Context Awareness for Recipes), which aims to enhance context-aware recipe following for people with vision impairments through tracking object status change that enables more accurate recipe progress tracking and provides contextual feedback. OSCAR extracts object status information, aligns cooking video/image data to recipe steps, and provides contextual support through sequential prediction and a time-causal model. By capturing the dynamic transformations of ingredients and maintaining context throughout the cooking process, OSCAR aims to offer higher granularity and accuracy for tracking cooking steps. Here are the research questions: (\textbf{RQ1}) How feasible is it to extract information of object status in cooking? (\textbf{RQ2}) How useful object status can enhance the performance of cooking progress prediction through VLMs? (\textbf{RQ3}) How could object status information be used to refer to the context of the cooking stages?

This paper presents OSCAR's design, implementation, and evaluation, focusing on recipe content processing, object status extraction, visual data alignment, and prediction. For this late breaking work, we only evaluated recipe tracking functionality by using the YouCook2 dataset \cite{ZhXuCoAAAI18} and our real-world non-visual cooking video dataset. Results show OSCAR's feasibility in tracking recipe steps with visual information with higher accuracy, showing more than a 20\% improvement in performance over the baseline for different VLMs. Our contributions include:

\begin{itemize}
\item Designing OSCAR to integrate recipe processing, object status extraction, visual tracking, and step prediction.
\item Demonstrating the utility of object status in enhancing recipe step tracking.
\item Identifying factors that impact recipe following prediction performance in real-world cooking scenarios.
\item Providing a new non-visual cooking video dataset with step annotations.
\end{itemize}

\section{Design and Methodology of OSCAR}
\begin{figure*}[]
    \centering
    \includegraphics[width=1\columnwidth]{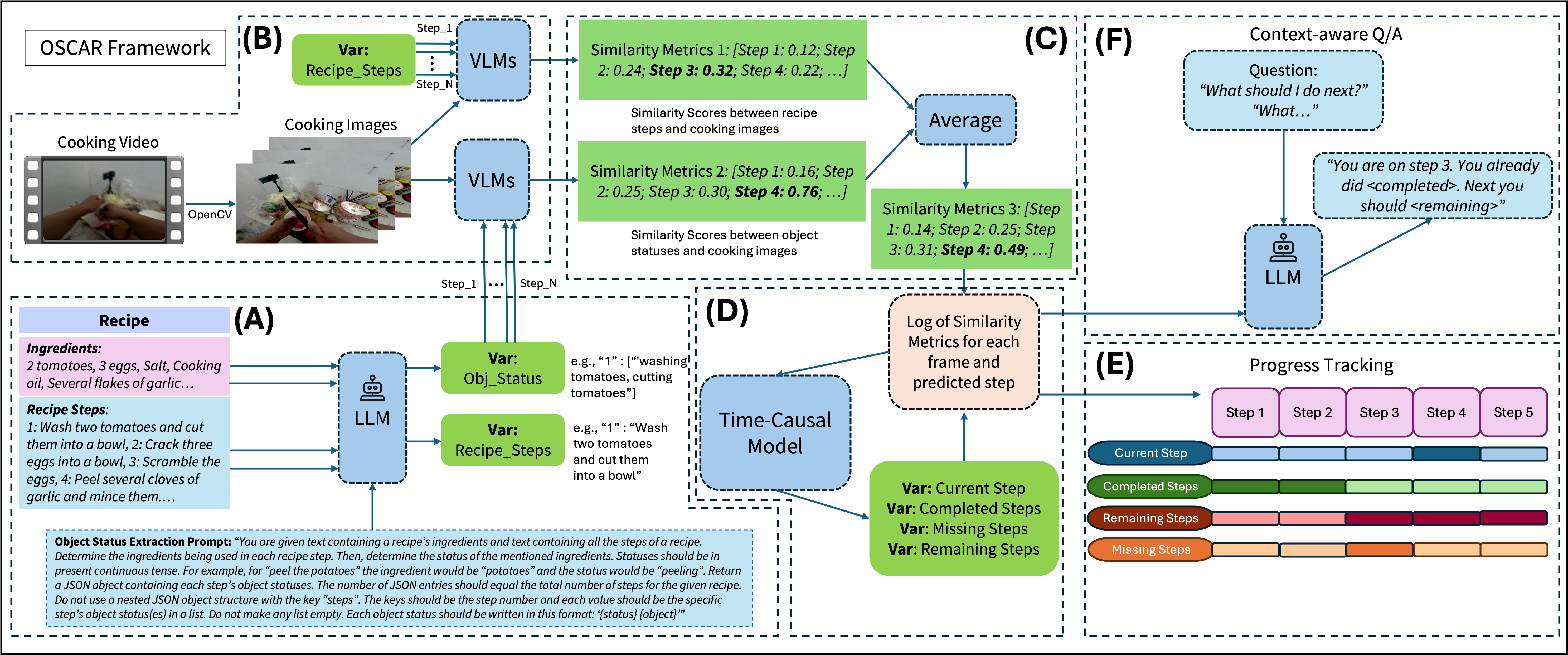}
    \caption{Illustration of OSCAR, application of progress tracking, and context-aware Q/A. (A) Recipe Formatting and Object Status Extraction, (B) Visual Data Extraction and Recipe Step Alignment, (C) Log of Similarity Metrics and Sequential Predictions, (D) Time-Causal Model, (E) Progress Tracking, (F) Context-aware Q/A.}
    \label{fig:tracking}
    \Description{This is a figure that shows the pipeline of OSCAR and application of progress tracking and context-aware Q/A. In the figure, it is a flow chart that shows OSCAR starts from extracting cooking video to cooking images, and align cooking images with object status and recipe steps that were extracted from original recipe and LLM by using VLMs. It then extracts the similarity metrics of alignment and saves it inside a log file that stores all predicted steps. It also passes the data through a time-causal model and extracts completed steps, completed steps, missing steps, and remaining steps, and saves that back in the log file. It then generates the progress tracking that includes the current step, completed steps, remaining steps, and missing steps. The log file is also used to parse to LLM that people can ask contextual questions and get feedback based on the history log.}
\end{figure*}

OSCAR (Object Status Context Awareness for Recipes) is a pipeline designed to support blind cooks by enabling context-aware recipe tracking (Figure \ref{fig:tracking}). It leverages object status tracking and conversational querying to enhance cooking guidance. OSCAR uses a combination of object status information from recipes and visual data (e.g., photos and videos) to recognize and track each step of a recipe. By utilizing state-of-the-art (SOTA) Vision-Language Models (VLMs) and Large Language Models (LLMs), OSCAR aligns recipe steps with visual frames, providing real-time contextual information and support throughout the cooking process. Here are the details about the key features of the pipeline:

\textbf{Recipe Formatting and Object Status Extraction}: Recipes often vary in formatting, which makes extracting step information challenging \cite{li2024recipe}. OSCAR uses GPT-4o \cite{achiam2023gpt} to standardize recipe steps, which are then saved as variables (Figure \ref{fig:tracking}). It extracts object status information by analyzing the ingredient list alongside the cooking actions \cite{xue2024learning}. The extracted statuses, such as ``chopping carrots'' or ``sautéing mushrooms,'' are stored in a structured format and linked to their respective recipe steps (Figure \ref{fig:tracking}).

\textbf{Visual Data Extraction and Recipe Step Alignment}: OSCAR processes visual data using OpenCV \cite{bradski2000opencv} to extract relevant frames from videos (Figure \ref{fig:tracking}). These frames are then aligned with recipe steps and object statuses using VLMs. The alignment ensures visual content corresponds accurately to the cooking process, reducing confusion and errors. By comparing similarity scores between visual frames and object statuses, OSCAR tracks recipe progress and averages similarity metrics to improve prediction accuracy \cite{luo2021clip4clip}.

\textbf{Log of Similarity Metrics and Sequential Predictions}: OSCAR maintains a detailed prediction log for each visual frame, which includes similarity scores, predicted steps, completed steps, and remaining steps (Figure \ref{fig:tracking}). This log supports progress tracking and allows users to ask contextual questions such as ``What step am I in?'' or ``Have I sautéed the mushrooms?'' OSCAR provides responses based on the prediction log (Figure \ref{fig:tracking}).

\textbf{Time-Causal Model}: To ensure sequential consistency, OSCAR uses a time-causal model \cite{stephan2020time,prabhakar2010temporal,pearl2018book} to enhance temporal coherence in predictions (Figure \ref{fig:tracking}). This model enforces logical step progression, reducing false positives and avoiding redundant or out-of-order predictions (Figure \ref{fig:tracking}). By doing so, OSCAR maintains the natural flow of cooking tasks, making the cooking experience more intuitive for users.

\textbf{Progress Tracking}: Based on the time-causal model and log of similarity metrics for each predicted step, OSCAR could perform recipe progress tracking of the current step, the completed steps, the remaining steps, and the missing steps (Figure \ref{fig:tracking}).

\textbf{Context-aware Q/A}: Based on the recipe content, the current image frame, and the log of cooking progress as prompts, OSCAR is able to provide contextual feedback through LLMs, such as what is the next step, or missing steps (Figure \ref{fig:tracking}). The demonstration is in the video figure. However, we did not evaluate this in this LBW.

\section{Evaluation with YouTube Cooking Instruction Video}
To evaluate OSCAR's performance in tracking and predicting recipe steps, we first tested it using large-scale recipe instruction video datasets with annotated timestamps and recipe steps. This evaluation aimed to measure OSCAR's effectiveness in aligning recipe steps with visual content through object statuses.

\subsection{Dataset}
We used the YouCook2 dataset \cite{ZhXuCoAAAI18} to validate OSCAR's ability to predict recipe steps using object statuses. YouCook2 includes 176 hours of untrimmed cooking videos, each averaging 5.27 minutes \cite{ZhXuCoAAAI18}. Videos are temporally annotated with start and end times for each procedural step, and contain diverse metadata such as segment duration and recipe step counts \cite{ZhXuCoAAAI18}. This variety makes YouCook2 suitable for assessing OSCAR's ability to handle complex cooking scenarios. Since YouCook2 did not include ingredient metadata, we manually reviewed videos to extract ingredient details from video content or publisher notes. Irrelevant videos, such as those featuring chef discussions, were filtered out. The final dataset consisted of 173 videos, with an average of 7.7 recipe steps per video.

\begin{table*}[]
\resizebox{1\columnwidth}{!}{%
\begin{tabular}{l|lllll}
VLM Model & Baseline Accuracy & Baseline Standard Deviation & OSCAR Accuracy & OSCAR Standard Deviation & Improvements \\ \hline
CLIP      & 41.7\%            & 17.5\%                      & 68.0\%         & 19.0\%                   & 26.3\%       \\
SigLIP    & 62.2\%            & 18.0\%                      & 82.8\%         & 14.7\%                   & 20.6\%       \\ \hline
\end{tabular}%
}
\caption{Evaluation results with recipe instruction videos.}
\label{table:youtubetesting}
\end{table*}

\subsection{Method}
\subsubsection{Data Processing and Baseline Prediction}
Videos were annotated with timestamps for each recipe step. Each video segment was divided into five equal parts, and a frame was randomly selected from each segment \cite{tong2022videomae}. We applied a blur filter to select the least blurry frame \cite{bradski2000opencv}. We then used CLIP \cite{radford2021learning} and SigLIP \cite{zhai2023sigmoid} to calculate similarity scores between frames and recipe steps as baseline. Predictions were repeated three times, and average accuracy was recorded. The decision to repeat the prediction process three times per step helps account for variability across frames within each step, especially in dynamic cooking scenes where objects or hands may occlude key items. Averaging accuracy across these repetitions provides a more reliable estimate, reducing the impact of transient visual noise such as motion blur or occlusion, which is common in cooking \cite{shi2021temporal}.

\subsubsection{Evaluating OSCAR for Object Status}
Object statuses were extracted from the ingredient list and corresponding cooking steps. Using the same frames as in the baseline method, similarity scores were calculated between frames and object statuses using CLIP and SigLIP. The time-causal model was then applied to further enhance performance. Final accuracy was obtained by averaging similarity scores across the frames.


\subsection{Results}
\subsubsection{Baseline}
Among 173 annotated instructional videos, the baseline accuracy was 41.7\% (SD = 17.5\%) for CLIP and 62.2\% (SD = 18.0\%) for SigLIP (Table \ref{table:youtubetesting}). These results highlight the challenges of accurately predicting recipe steps, especially where visual cues are ambiguous.

\subsubsection{OSCAR}
With object status information and a time-causal algorithm, OSCAR significantly outperformed the baseline. CLIP's accuracy improved to 68.0\% (SD = 19.0\%), and SigLIP's to 82.8\% (SD = 14.7\%) (Table \ref{table:youtubetesting}). This demonstrates the value of object status in improving accuracy, especially for complex cooking scenarios. OSCAR's performance improvements (26.3\% for CLIP, 20.6\% for SigLIP) resulted from its ability to provide visual context for ambiguous steps. For instance, OSCAR helped differentiate between similar actions, such as adding ingredients twice during different phases of a recipe, by incorporating a time-causal model.

\begin{figure}[]
    \centering
    \includegraphics[width=1\columnwidth]{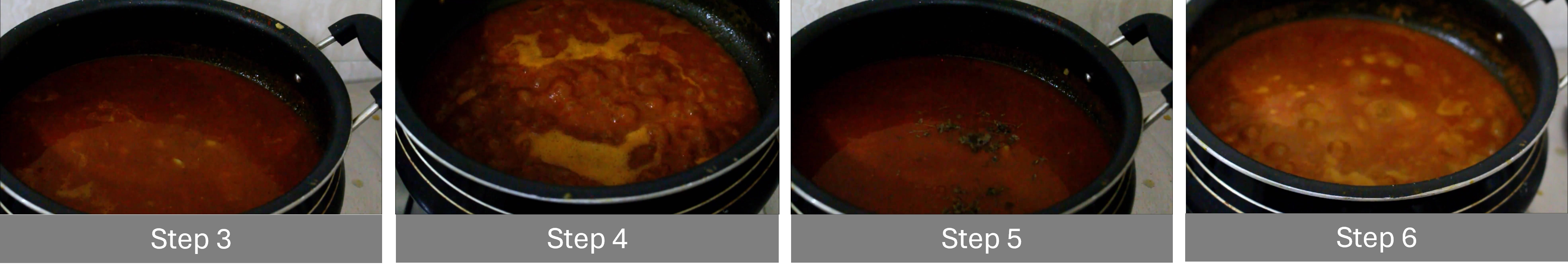}
    \caption{Image frames of cooking video that show similar visual frames of different steps that caused misprediction.}
    \label{fig:dishtype}
    \Description{This figure includes four subfigures marked from step 3 to step 6. They all show a pot of boiling soup inside. In steps 4 and 6, we can see that there might be some beans inside, and there are some possible herbs in step 5.}
\end{figure}

\subsection{Factors Impacting Performance}
We analyzed factors affecting the performance of visually tracking recipe steps, including exocentric views, subtle visual changes, atypical ingredient descriptions, and low-quality video.

\textbf{Exocentric vs. Egocentric View}: Performance dropped when videos showed exocentric views, displaying the entire kitchen environment rather than focusing on cooking actions. Obstructions, such as tools blocking ingredients, also hindered VLMs from accurately associating frames with recipe steps. \textbf{Subtle Visual Changes}: Dishes with gradual changes, like soups, posed challenges due to minimal visual differences between steps. This led to difficulty in aligning cooking steps with recipe instructions. \textbf{Unmatched Descriptions}: VLMs struggled with unusual ingredient descriptions not matching typical training data. For example, in a Korean Watermelon Rind Kimchi recipe, CLIP failed to accurately identify the rind due to its unique visual characteristics. \textbf{Low-Quality Video}: Low-resolution videos and visual impacts, like steam blurring the view, decreased prediction accuracy. Such issues obscured important cues needed for tracking the cooking process.


\begin{table*}[t]
\resizebox{1\columnwidth}{!}{%
\begin{tabular}{l|lllll}
VLM Model & Baseline Accuracy & Baseline Standard Deviation & OSCAR Accuracy & OSCAR Standard Deviation & Improvements \\ \hline
CLIP      & 33.7\%            & 21.8\%                      & 58.4\%         & 16.5\%                   & 24.7\%       \\
SigLIP    & 41.9\%            & 14.0\%                      & 66.7\%         & 18.7\%                   & 24.8\%       \\ \hline
\end{tabular}%
}
\caption{Evaluation results with non-visual cooking dataset.}
\label{table:nvcheftesting}
\end{table*}

\section{Evaluation with Real-world Cooking Dataset from People with Vision Impairments}
To assess OSCAR's effectiveness in real-world non-visual cooking, we collected a dataset of 12 cooking videos featuring individuals with vision impairments cooking naturally in their own kitchens and evaluated the performance of OSCAR.

\begin{figure}[t]
    \centering
    \includegraphics[width=\columnwidth]{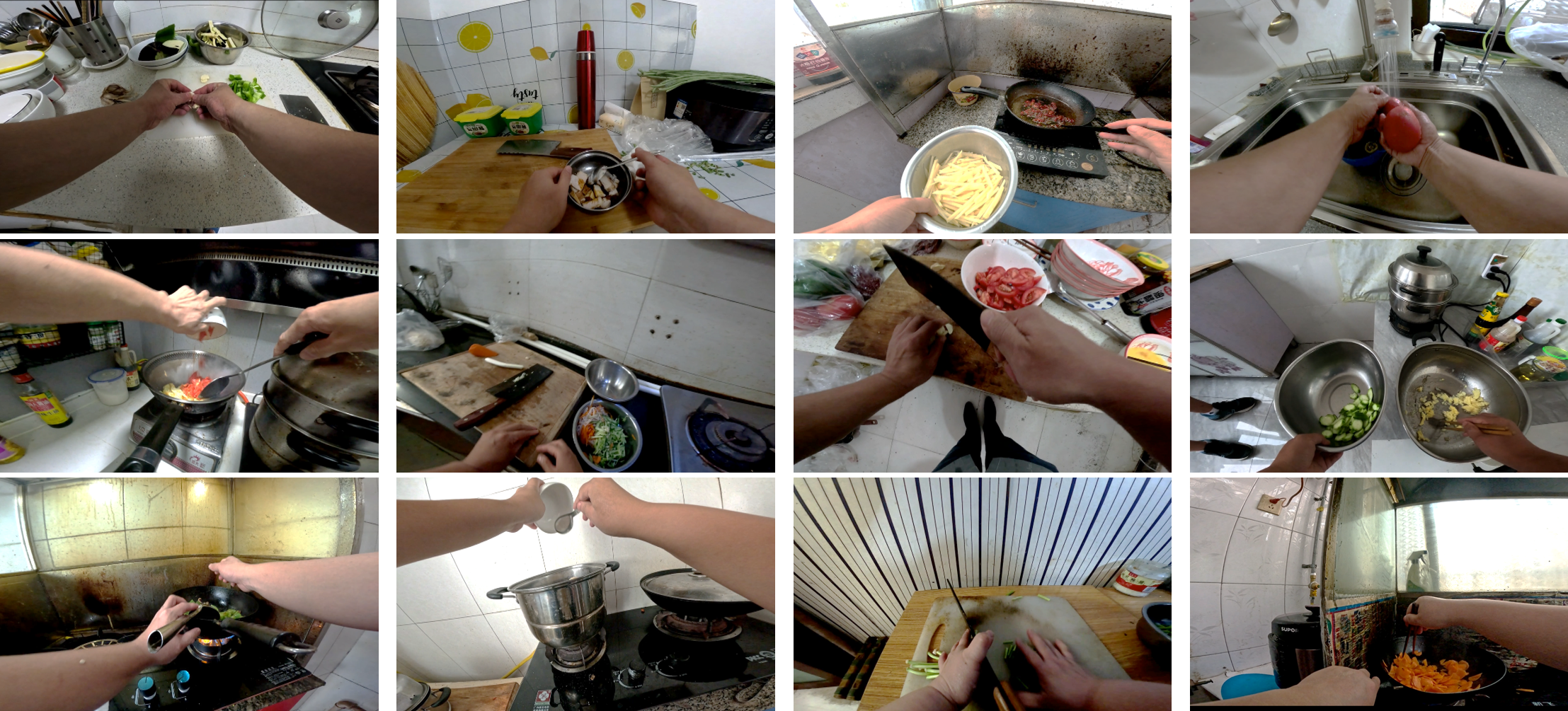}
    \caption{Thumbnail of the non-Visual cooking dataset of 12 videos by people with vision impairments.}
    \label{fig:thumbnail}
    \Description{The image displays a grid of twelve scenes (3 x 4) showing a person's hands performing various cooking tasks from a first-person perspective. These tasks include chopping vegetables, washing ingredients, stirring food in a pan, and mixing ingredients in bowls.}
\end{figure}

\subsection{Non-visual Cooking Dataset Characteristics}
To provide the benchmark for evaluating different VLMs for non-visual cooking, we first collected a video dataset that includes 12 cooking videos (V1–V12) of individuals with vision impairments (6 male, 6 female, average age of 43) preparing dishes in their own kitchens. Video lengths range from 14 to 51 minutes, with an average of 8.1 steps per recipe. Four of them are legally blind and eight are totally blind. Recorded in 5.2K resolution using a GoPro 11 on a chest band through an egocentric view, the videos were down-sampled to 1080P for analysis. Following YouCook2 \cite{ZhXuCoAAAI18}, we annotated timestamps for each recipe step, aligning them with visual content. For instance, one video details washing, chopping, and assembling a salad. Participants were compensated with \$40 USD. The data collection was approved by our Institutional Review Board (IRB). Here is the dataset link: \url{https://huggingface.co/datasets/limingz3/Non-visual_Cooking_Video_Dataset}

\subsection{Evaluation Methods} 

\subsubsection{Data Processing and Baseline Prediction}
Each video step was divided into five equal segments, with one frame randomly selected per segment \cite{tong2022videomae}. A blur filter was applied to select the least blurry adjacent frame \cite{bradski2000opencv}. Using CLIP \cite{radford2021learning} and SigLIP \cite{zhai2023sigmoid}, similarity scores between frames and recipe steps were calculated. The highest average similarity score determined the predicted step. Accuracy was marked as 100\% if the prediction matched the ground truth; otherwise, 0\%. This process was repeated three times per step, and the average accuracy was used as the final score for each step, repeated across all recipe steps and videos. The prediction process was repeated three times per step to reduce the impact of frame-level variability and ensure more stable accuracy estimates, following multi-sampling strategies used in video analysis tasks \cite{shi2021temporal}.

\subsubsection{Evaluating OSCAR for Object Status}
Object status information was extracted from ingredient lists and cooking steps. Using the same frames as in the baseline method, similarity scores between frames and object statuses were calculated with CLIP and SigLIP. A time-causal model was applied to improve performance. Similarity scores were averaged across frames and repetitions, then combined with baseline scores to determine final prediction accuracy, highlighting the added value of object statuses compared to text-only predictions \cite{luo2021clip4clip}.

\subsection{Results}

\subsubsection{Baseline}
The baseline results revealed an average accuracy of 33.7\% (SD = 21.8\%) for recipe step predictions using CLIP (Table \ref{table:nvcheftesting}). SigLIP, on the other hand, achieved an average accuracy of 41.9\% (SD = 14.0\%) (Table \ref{table:nvcheftesting}). These results highlight the challenges of accurately aligning recipe steps with image frames.


\subsubsection{OSCAR}
By incorporating object status information and utilizing a time-causal algorithm, OSCAR improved the prediction accuracy. The average accuracy for CLIP increased from 33.7\% to 58.4\% (SD = 16.5\%) (Table \ref{table:nvcheftesting}), while SigLIP’s performance rose from 41.9\% to 66.7\% (SD = 18.7\%) (Table \ref{table:nvcheftesting}). These improvements demonstrate the effectiveness of using object statuses to enhance OSCAR's ability to accurately track and predict recipe steps, especially in complex cooking scenarios. Our analysis revealed significant performance improvements for both CLIP (24.7\%) and SigLIP (24.8\%) (Table \ref{table:nvcheftesting}), highlighting the importance of object status recognition in non-visual cooking. OSCAR with SigLIP achieved 100\% accuracy for V12, improving baseline performance by 60\%. 

Evaluating vision models with real-world cooking data is crucial to accurately track cooking progress. In the dataset, cooks with vision impairments often take longer to locate and interact with objects, leading to potential false positives when similar items are involved, as VLMs may misinterpret prolonged handling as an action. Prioritizing object status recognition before aligning steps reduces such errors, enhancing tracking accuracy. Additionally, people with vision impairments frequently use personalized tools that differ from those in conventional recipes \cite{li2024recipe,li2024contextual,li2021non}. Aligning only recipe steps with frames can result in poor performance, as tools in recipes often differ from those used. Object status, however, remains consistent. Focusing on object status improves OSCAR’s tracking accuracy, minimizes the need for users to adapt to unfamiliar tools, and supports a more inclusive, flexible system adaptable to diverse cooking environments while preserving user autonomy.

\begin{figure}[]
    \centering
    \includegraphics[width=1\columnwidth]{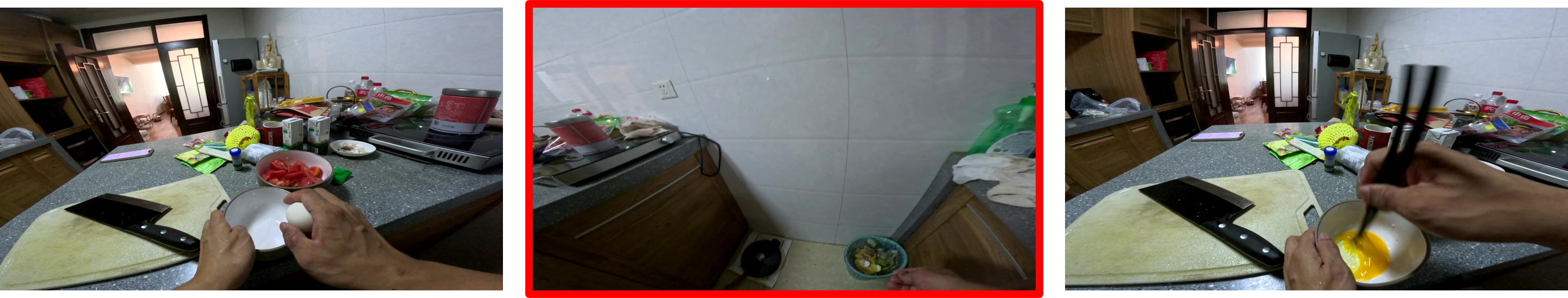}
    \caption{These three frames were captured for V4 during step 1: `Crack an egg and scramble it.' The middle frame showed the data where the blind cook was looking for a garbage bin to throw the egg shell, which got captured and impacted the performance of predicting the step.}
    \label{fig:lookingforobjects}
    \Description{The image presents a sequence of three first-person cooking scenes. In the first scene, the individual is handling a bowl of chopped tomatoes on a cutting board, with a cleaver resting nearby on a cluttered kitchen countertop, while the living room is visible in the background. In the second scene, the person appears to be discarding food waste, holding a small bowl near a garbage bin, with various cooking items and packaging visible on the counter. In the third scene, the individual is whisking eggs with chopsticks in a small bowl, continuing the cooking process, with the same kitchen countertop and supplies in the background.}
\end{figure}

\subsection{Factors Impacted the Performance and Design Implications}

We analyzed factors impacting VLM performance in real-world scenarios involving people with vision impairments. Two researchers collaboratively examined cases of underperformance, comparing object status descriptions with video frames to identify key findings.

\subsubsection{Impact of Implicit Tasks on Recipe Progress Prediction} In V4, the prediction accuracy dropped to 46.7\% (compared to the average of 66.7\%) due to implicit tasks like disposing of eggshells during the step "crack an egg and scramble it." (Figure \ref{fig:lookingforobjects}) These unlisted but common actions, such as cleaning or organizing tools, often caused mispredictions. For individuals with vision impairments, these tasks take longer, increasing the likelihood of errors in step alignment.

\subsubsection{Rechecking and Securing Tools}
Participants rechecked tools and ingredients frequently before and after use, as observed in video V3, where the cook repeatedly revisited chopped tomatoes before they were needed. This behavior, essential for accuracy and accessibility, confused VLMs, which misinterpreted such actions as new recipe steps. Designing models to differentiate these checks from actual tasks is crucial for progress tracking.

\subsubsection{Lighting Conditions} Lighting inconsistencies affected object detection accuracy. High-contrast lighting caused VLMs to focus on the background, while dim lighting obscured ingredients. For example, V6 had only 40\% accuracy due to poor lighting, compared to the average of 66.7\%. Future systems should incorporate real-time lighting assessment and guidance to improve usability. 

\subsubsection{Camera Angles and Field of View} Improper camera angles led to distorted or missing objects in the frame, reducing prediction accuracy. For instance, in V6, a high-angle camera misaligned with the workspace resulted in slightly lower performance. Future solutions should help users optimize camera positioning non-visually.

\section{Discussion}

\subsection{Real-World Non-visual Cooking}
We found that OSCAR performed better on the YouCook2 dataset than on the real-world non-visual cooking dataset, as YouCook2 features edited videos with close-up shots, consistent lighting, and professional angles. In contrast, real-world conditions, such as in V6 (40\% vs. 66.7\% average), introduce challenges like poor lighting and limited camera views, leading to more false positives. Previous research on cooking progress prediction often relies on streamlined datasets, overlooking the complexity of real-world cooking, especially for individuals with visual impairments, whose interactions with ingredients and tools vary widely (e.g., \cite{lin2020recipe, cao2019video, gong2021temporal, jiao2021new, ciaparrone2020deep, yao2020video}).  Furthermore, we demonstrated the usefulness of OSCAR using CLIP and SigLIP via the Hugging Face API \cite{huggingface2024clip}, which completes each call in under 1000 milliseconds for tracking cooking steps. However, future research exploring the use of more comprehensive Vision-Language Models (VLMs) should carefully evaluate and validate the impact of increased processing time on both step tracking and the delivery of timely feedback to end users. While OSCAR shows promise in both controlled and real-world settings, further advancements are needed to make VLMs more effective for non-visual cooking scenarios.

\subsection{Enhancing Information Granularity on Object Statuses and Other information} 
OSCAR demonstrates its ability to align object statuses with VLMs to improve accuracy in step prediction. Advances in VLMs and computational power could improve the granularity and precision of these predictions, enabling finer tracking of actions and object statuses. For example, OSCAR might distinguish stages of chopping or stirring and detect subtle changes in ingredient status, such as cooking times or textures. These enhancements could provide more accurate and granular guidance for non-visual cooking. Furthermore, to address minimal visual changes that may reduce detection accuracy, future research could incorporate more granular data beyond object statuses, such as capturing user actions at each step, to enhance detection accuracy and compensate for the subtle visual changes in certain dishes, such as soups. As VLMs evolve, integrating these capabilities into OSCAR could empower people with vision impairments to cook more independently and confidently at home or in educational settings \cite{li2024recipe}.

\subsection{Expanding the Non-Visual Cooking Dataset} 
Our study contributes to Accessibility and Computer Vision by developing the first non-visual cooking dataset. It includes recordings of 12 participants with vision impairments cooking in their own kitchens, annotated with timestamps and corresponding recipe steps. This unique dataset captures natural cooking behaviors of blind cooks, enabling researchers to evaluate vision models and understand cooking progress in this context. Unlike generic cooking datasets, it highlights the specific challenges and needs of non-visual cooking. In future research, further data collection with diverse tasks, environments, and participants is needed to refine models and ensure assistive technologies can effectively support a broader range of individuals with vision impairments.

\subsection{Next Step: Accessible User Interface, Interaction, and User Evaluation}
In this paper, we used a real-world non-visual cooking dataset to evaluate OSCAR's performance in recipe following. However, we did not explicitly explore or study the interaction between the end user and OSCAR through user studies regarding providing contextual feedback. In our next step, we will focus on developing an accessible mobile prototype and conducting in-situ studies to better understand user input methods, conversational agents, and interaction preferences. This would provide valuable insights into how users engage with OSCAR and inform improvements for more effective and intuitive user interaction. We will then conduct in-situ user evaluation with blind cooks in their own kitchens.

\subsection{Limitation and Future Work}
In our research, our real-world non-visual cooking dataset only contained 12 videos from one single country. We believe that evaluating OSCAR with more real-world cooking datasets or samples from different cultures would further validate the usefulness of object statuses in recipe step alignment and how the VLMs could be improved to incorporate different environments, cooking items, and styles. Also, we only evaluated the usefulness of object statuses for recipe following with two SOTA VLMs (i.e., CLIP and SigLIP), future research could conduct more evaluations with other VLMs or Large Multimodal Models (e.g., GPT-4o) to enhance the generalizability of OSCAR in recipe tracking with object statuses, and we believe the performance will get improved with more powerful models. Moreover, given that existing vision models all have different performances, it is important to further explore how good is enough and create full benchmark based on user experiences.

\section{Conclusion}
In this paper, we introduced OSCAR (Object Status Context Awareness for Recipes), a novel approach to providing context-aware cooking support for individuals with vision impairments. OSCAR leverages VLMs to extract object status information and align it with recipe steps, offering contextual guidance that is crucial for independent cooking. Through evaluations with both a large-scale cooking video dataset and a real-world non-visual cooking dataset by people with vision impairments, we demonstrated OSCAR’s capability of leveraging object status tracking and time-causal model to significantly improve recipe tracking performance upon baseline models. Our findings highlight the potential of OSCAR to track more granular cooking steps and provide contextual feedback. We also presented design implications and factors that impact the prediction performance for real-world cooking scenarios (e.g., impact of implicit tasks on recipe progress prediction). Additionally, the release of the non-visual cooking dataset sets a new benchmark for developing assistive technologies, paving the way for future research and advancements in accessible cooking technologies for people with vision impairments.


\bibliographystyle{ACM-Reference-Format}
\bibliography{main}


\end{document}